\documentstyle[aps,prl,multicol,epsfig]{revtex}
\begin{document}
\draft
\title{Coupled Microwave Billiards as a Model for Symmetry 
       Breaking}
\author{H. Alt$^{1}$, C. I. Barbosa$^{2}$, H.-D. Gr\"af$^{1}$, T. Guhr$^{2}$,
        H.L. Harney$^{2}$,\\
        R. Hofferbert$^{1}$, H. Rehfeld$^{1}$, and A. Richter$^{1}$\\
       }
\address{$^{1}$
         Institut f\"ur Kernphysik, Technische Universit\"at Darmstadt,
         D-64289 Darmstadt, Germany\\
         $^{2}$Max-Planck-Institut f\"ur Kernphysik, D-69029 Heidelberg, 
         Germany\\        
        }
\date{\today}
\maketitle              
\begin{abstract}
Two superconducting microwave billiards have been electromagnetically coupled 
in a variable way. The spectrum of the entire system has been measured and the 
spectral statistics analyzed as a function of the coupling strength. It is 
shown that the results can be understood in terms of a random matrix model of 
quantum mechanical symmetry breaking -- as e.g. the violation of parity or 
isospin in nuclear physics.
\end{abstract} 
\begin{multicols}{2}
\narrowtext                                                   
Both, classical and quantum mechanical chaos can be studied with the help of 
billiards, see e.g. the recent review \cite{Report}. Quantum mechanical 
billiards are readily simulated by sufficiently flat microwave resonators 
\cite{Stoeckmann,Smilansky,Sridhar,Graef} since the Schr\"odinger and the 
Helmholtz equations are equivalent in two dimensions. Here, we present a study 
of a system consisting of two coupled resonators. This simulates the breaking 
of a symmetry, see below. The resonators were made superconducting. This
allowed us to study the transition from the uncoupled case into the regime of 
weak coupling very precisely. The positions of the resonances were determined 
with a precision of $10^{-7}$.

Symmetry breaking in chaotic systems has been intensely investigated.
Impressive is the study of parity violation in heavy nuclei \cite{Triple}.
Atomic and molecular symmetries were studied in \cite{Rosenzweig,Cederbaum}.
Another example from nuclear physics is isospin mixing, see e.g. 
\cite{Harney86,Mitchell}. In \cite{Mitchell}, the complete spectrum of the 
nucleus $^{26}$Al at low excitation energy was established. The analysis of 
this spectrum in terms of the so called Gaussian Orthogonal Ensemble (GOE) of 
random matrices showed that the level statistics was intermediate between a 
2-GOE and a 1-GOE behavior \cite{Guhr}. By this, we refer to the following 
model. Each level in the spectrum of $^{26}{\rm Al}$ can be characterized by 
isospin 0 or 1. In the absence of mixing, the spectrum of the states of each 
isospin (i.e. of each symmetry class) has the statistical properties of the 
eigenvalues of matrices belonging to the GOE. The superposition of the two 
spectra displays a 2-GOE behavior. It is described by the first term of the 
Hamiltonian
\begin {eqnarray}
  {\cal H}= \left( \begin{array}{cc} \framebox[1.1cm]{\rule[-1.5mm]{0cm}
                      {0.5cm}$GOE$} & 0\\
                        0 & \framebox[1.1cm]{\rule[-1.5mm]{0cm}
                      {0.5cm}$GOE$}
                        \end{array}\right) +  \alpha
               \left( \begin{array}{cc} \framebox[1cm]{\rule[-1.5mm]{0cm}
                      {0.5cm}0} &  V\\
                       V^+  &   \framebox[1cm]{\rule[-1.5mm]{0cm}{0.5cm}0}
                         \end{array}\right).  \label{Block}
\end {eqnarray}
This is a special case of the model of Ref. \cite{Rosenzweig}. The 
off-diagonal matrix couples the classes. The random elements in the $GOE$s and 
in $V$, all have the same rms value $v$ so that $\alpha=1$ makes ${\cal H}$ as 
a whole to be a GOE matrix. The resulting spectrum displays 1-GOE behavior. 
For the observables studied below, the 1-GOE behavior is actually reached 
already if $\alpha v/D$ is $\approx 1$. Here, $D$ is the mean level distance 
of ${\cal H}$. For simplicity, we set $v=1$ in the sequel. This makes $D$ 
dimensionless and the parameter governing the level statistics is then 
$\alpha/D$. -- In the example of $^{26}{\rm Al}$, this parameter was determined
from the level statistics; whence the mean square Coulomb matrix element that 
breaks isospin was derived. The present experiment tests the model 
(\ref{Block}) with a large number of levels and very clean spectra in a 
situation, where the parameter $\alpha/D$ controlling the symmetry breaking 
can be varied. Alternative models 
for coupled chaotic systems are e.g. given in \cite{botoul,ditschko}.

\vspace*{-0.375cm}
\begin{figure} [hbt]
\centerline{\epsfxsize=8.6cm
\epsfbox{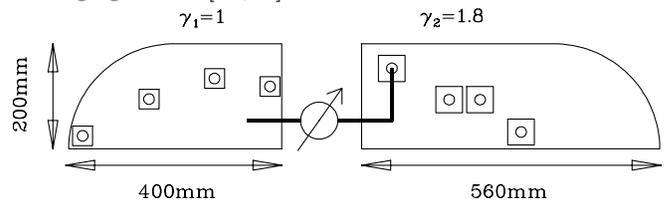}}
\caption{Shapes and locations of the antennas of the two coupled Bunimovich 
stadium billiards.}\label{logo}
\end{figure}
\vspace*{-0.525cm}

\begin{figure} [hbt]
\centerline{\epsfxsize=8.6cm
\epsfbox{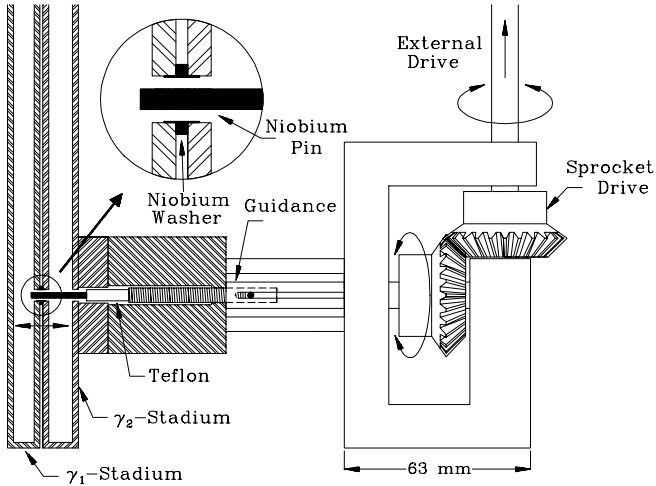}}
\caption{The adjustable superconducting coupling.}\label{wmd}
\end{figure}

In the present experiment each of the two symmetry classes consists of the 
eigenstates of a (quarter of a) stadium billiard, see Fig.~\ref{logo}. The 
radius of the quarter circle was in both cases $r=0.2$ m. The ratios $\gamma$ 
between the length of the rectangular part and $r$ were $\gamma_1=1$ and 
$\gamma_2=1.8$ for the two billiards, respectively.
The measurement was restricted to frequencies below 16 GHz where both 
resonators are two-dimensional and display 608 and 883 resonances, 
respectively. For the variable superconducting 
coupling, the two resonators were put on top of each other and holes, 4 mm in 
diameter, were drilled through the 2 mm thick walls of both resonators (see 
Fig.~\ref{wmd}). A niobium washer ensured sufficient electrical contact 
between the resonators. Coupling was achieved through a niobium pin, 2 mm in
diameter, which could be moved perpendicularly to the plane of 
the billiards from outside the helium cryostat by a drive. The coupling
strength is determined by the depths $x_1$ and $x_2$ by which the niobium pin
penetrates into the $\gamma_1$- and $\gamma_2$-stadium, respectively. For the
strongest coupling, a second niobium pin, penetrating all the way through both
resonators, was added. Stronger coupling could have been obtained by using 
even more coupling pins. This was, however, not realized since it was the 
particular emphasis of the present experiment to study the transition from the 
uncoupled case into the weakly coupled regime.

\vspace*{-0.38cm}
\vspace*{-0.6cm}

For each of the couplings we have measured the complete spectrum in steps of 
10 kHz. In doing so, half of the microwave power was fed into each of the 
resonators. The states of the $\gamma_1$-stadium were not always visible 
through an antenna connected to the $\gamma_2$-stadium and vice versa. 
Therefore the spectrum of the entire system was constructed by adding the 
spectra obtained through the 7 antennas, four on the $\gamma_1$- and three on 
the $\gamma_2$-stadium, see Fig.~\ref{logo}.

\begin{figure} [hbt]
\centerline{\epsfxsize=8.6cm
\epsfbox{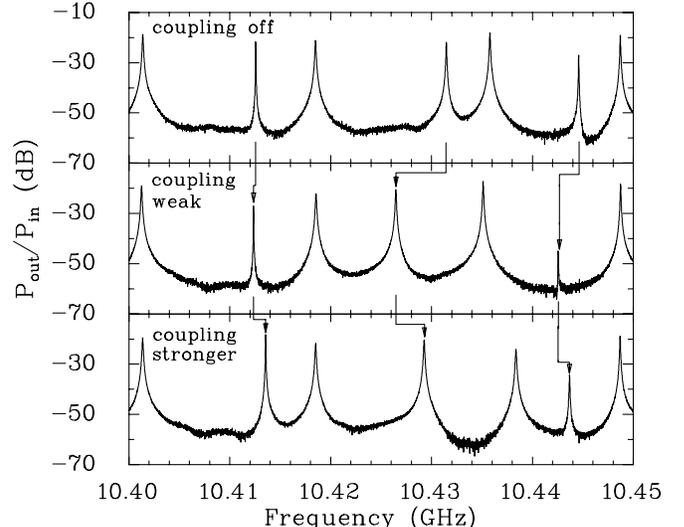}}
\caption{A small frequency range of three spectra with different coupling.
 The arrows are intended to help to recognize the shifts of a few resonances.}
\label{Spektrum}
\end{figure}

A small frequency range of spectra at various couplings is shown in 
Fig.~\ref{Spektrum}. 
One recognizes that  the resonances are shifted by statistically varying 
amounts. The observed frequency shifts result from the coupling of
the two cavities and from the perturbation of the electromagnetic field by 
the pin. The latter effect was investigated separately by inserting (or not) 
the pin into the $\gamma_2$-stadium only. Under that condition when only the 
field is perturbed but no coupling is achieved all measured spectra (see e.g. 
top of Fig.~4) displayed a clean 2-GOE behavior.
The mean level spacing is $D=10.7$ MHz. The resonance widths are of the order 
of 1 to 100 kHz. The high $Q$ of $10^5 - 10^6$ together with the very good 
signal to noise ratio of up to 50 dB  of the superconducting setup was 
obviously necessary to detect the partly small shifts and thus the dependence 
of the level statistics on the coupling. 

We now turn to the analysis of the spectra. From the ansatz (\ref{Block}), the 
nearest neighbor spacing distribution (NND), the $\Sigma^2$-statistic and the 
$\Delta_3$-statistic can be obtained numerically, see \cite{Guhr,French} and 
results below. Although the coupling parameter $\alpha/D$ can be determined by 
comparing numerical simulations with the data, it is convenient to have 
analytical expressions of $\Sigma^2$ etc. as functions of $\alpha/D$. French 
et al. \cite {French} and Leitner et al. \cite{Leitner} have derived them for 
small coupling parameters using perturbation theory.

The analysis of the present data has been based on the $\Sigma^2$-statistic or 
number variance
\begin{equation}
  z(L) = \langle (n(L)-L)^2\rangle.
\end{equation}
Here, $n(L)$ is the number of eigenvalues in an interval of length $L$. To 
obtain $z(L)$, we divided the entire unfolded spectrum of length $ND$ into 
$N_L=N/L$ adjacent intervals of length $L$ and took the average 
$\langle \rangle$ over these. By looking at the correlation between 
$z(L)$ and $z(L+\varepsilon)$, we convinced ourselves that $z(L)$ and 
$z(L+\varepsilon)$ were statistically independent for $\epsilon \ge 0.025$
-- at least in the range $1  \le L \le 5 $. Calculating $z(L)$ in that range in
steps of $\varepsilon=0.025$ has provided $M=161$ experimental numbers 
$z(L_k)$, $k=1 \dots M$, that were statistically independent as is needed for 
the fit-procedure described below. The upper limit $L \le 5$ is defined by the 
saturation of the $\Sigma^2$-statistic \cite{Berry} : Random matrix theory is 
known to describe spectral fluctuations of chaotic systems up to a maximum
$L_{max}$ which is related to the length of the shortest periodic orbit. This
sets $L \le 5$ here. To check the influence of the saturation, we restricted 
the extraction of $\alpha/D$ to $L\le 3$. The slight change of the results was 
well within the errors. The lower limit of  $1\le L$ is explained in the sequel.

\vspace*{-0.35cm}
\vspace*{-0.5cm}

The expectation value $\overline{z(L)}$ is called $\Sigma^2(L,\Lambda)$, where 
$\Lambda=(\alpha/D)^2$. According to Ref. \cite{Leitner}, this function is 
\begin{equation}
  \Sigma^2(L,\Lambda)=\Sigma^2(L,\infty)+\frac{1}{\pi^2} \ln \left(
  1+\frac{\pi^2 L^2}{4(\tau +\pi^2\Lambda)^2}\right).\label {Sigma2}
\end{equation}
Here, $\Sigma^2(L,\infty)$ is the limiting function for the 1-GOE system. The 
parameter $\tau$ is related to the ratio of dimensions in the GOE blocks of 
eq. (\ref{Block}). One finds $\tau=0.74$ in the present case. 
  
In order to estimate $\alpha/D$, one has to know
the probability distribution $w_k(z(L_k) | \Lambda), k=1 \dots M$, of every
data point. By applying the ``bootstrap method'' \cite{bootstrap} to the set of
$N_{L_k}$ intervals from which $z(L_k)$ was calculated, we found $w_k$ to 
be a $\chi^2$-distribution with average value as given by (\ref{Sigma2}) and 
with roughly $N_{L_k}$ degrees of freedom -- which is reasonable. However, this
was true only for $L_k \ge 1$. For $L_k<1$, no analytical representation of 
the distribution of $z(L_k)$ was found. At the same time, the information on
$\Lambda $ is lost: 
for small $L $, the relative change of $\Sigma^2$ with $\Lambda$ is 
of the order of $L$ while the relative rms deviation of $z$ is of the order of 
$L^{1/2}$. Therefore, the analysis was restricted to $L_k \geq 1$.

\vspace*{-0.25cm}
\begin{figure} [hbt]
\centerline{\epsfxsize=8.6cm
\epsfbox{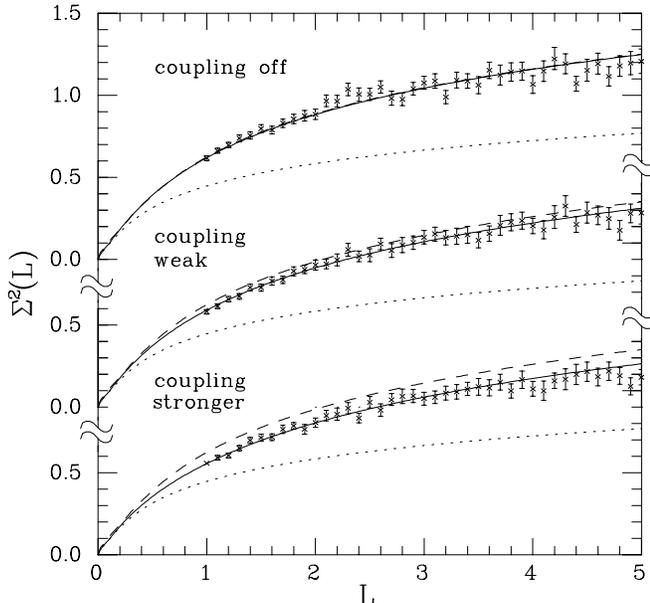}}
\caption{The $\Sigma^2$-statistic for three different couplings.  
 The dotted line gives the 1-GOE and the dashed line the
 2-GOE behavior. The solid lines result from the estimation of $\alpha/D$ 
 described in the text and correspond to the first, the third and the last entry 
 of Table~\ref{alpha_NND}.} 
\label{Sigma2bild}
\end{figure}
\vspace*{-0.5cm}

The joint distribution $W(z|\Lambda) = \prod_k w_k(z(L_k)|\Lambda)$ of all the 
$z(L_k)$ was converted into the distribution $W(\Lambda|z)$ of $\Lambda$ with 
the help of Bayes' theorem
\begin{equation}
  W(\Lambda|z) = \frac{W(z|\Lambda)\mu(\Lambda)}{\int d\Lambda 
  W(z|\Lambda)\mu(\Lambda)}. \label {Bayes}
\end{equation} 
The a priori distribution of $\Lambda$ was defined as
\begin{equation}
\mu(\Lambda)=\left|\int d^{M}z\; W(z|\Lambda) \frac{\partial^2}{\partial 
   \Lambda^2} \ln W (z|\Lambda)\right|^{1/2}
\end{equation}
since this expression ensures that -- at least for sufficiently large $M$ --
the entropy 
\begin{equation}
  H = - \int d^{M}z\; W(z|\Lambda) \ln\left(\frac{W(z|\Lambda)}{\int d\Lambda 
  W(z|\Lambda)\mu(\Lambda)}\right)
\end {equation}
of $W(z|\Lambda)$ is independent of $\Lambda$, whence $\Lambda$ cannot be 
estimated by a maximum entropy argument without any experiment. 

The center and the rms deviation of the distribution (\ref{Bayes}) were 
determined if it was Gaussian. This defines the best estimate and the error of 
$\Lambda$ as well as $\alpha$ given in Table~\ref{alpha_NND}. For three cases 
(part of) the data and the fit function (\ref{Sigma2}) that results from the 
best estimate are given in Fig.~(4). In the case of zero coupling, 
$P(\Lambda|z)$ was not a Gaussian and the result given in Table~\ref{alpha_NND}
is an upper limit for the confidence of $68\%$.

A $\chi^2$-test -- generalized to the case of non-Gaussian distributions 
$w_k$ -- has shown that for the fits reported in Table~ \ref{alpha_NND}, 
the analytical model (\ref{Sigma2}) of \cite{Leitner} is compatible with
the data. To check that the perturbation result (\ref{Sigma2}) does apply to 
our case, we also performed numerical simulations of the full model 
(\ref{Block}) using the procedure described in \cite{Guhr}. Thereby we obtained
values for $\alpha/D$ which cannot be affected by limitations of the 
perturbative calculation of \cite{Leitner}. However, reassuringly, the results
of both analyses are consistent within the errors.

From the coupling parameter and the mean level spacing $D=10.7$ MHz, one
obtains the root mean square coupling matrix element $\alpha $ of the model
(\ref{Block}). It corresponds to the rms Coulomb matrix element
$\sqrt{\langle H_C^2 \rangle}$ that is responsible for isospin mixing
\cite{Guhr}. Figure~\ref{Sigma2bild} shows that starting from 2-GOE behavior 
in the uncoupled case ($\alpha/D\le0.024$) one moves through the weakly 
coupled case ($\alpha/D=0.13$) towards 1-GOE behavior. The strongest coupling 
($\alpha/D=0.20$) realized here causes, however, still a relatively weak 
symmetry breaking of about the same size as the isospin symmetry breaking in 
$^{26}{\rm Al}$. The spreading width $\Gamma^\downarrow/D = 2\pi (\alpha/D)^2$
which is a measure of how much the states of the two symmetry classes are mixed
into each other, is also given in Table~\ref{alpha_NND}. In the case of the 
strongest coupling e.g. one sees that a state of class 1 carries about 
25\% admixture of class 2 and vice versa. This is the reason for the shifts 
observed in Fig.~\ref{Spektrum}.

The fact that the level statistics depend on $\alpha$ proves that the 
coupling block $V$ in the Hamiltonian (\ref{Block}) is essentially filled with 
statistically independent elements -- as we have assumed. If -- on the contrary
-- $V$ were a separable interaction coupling only one specific configuration of
the first symmetry class to one in the second class, then the level statistics
would not change as a function of $\alpha$.

\vspace*{-0.25cm}
\begin{table}[thb]
\caption {Mixing parameters for six different coupling strengths resulting
          from the Bayesian analysis described in the text. The penetration
          depths of the coupling pin into the resonators is given by
          $(x_1,x_2)$ in mm. The results of the bottom row were obtained 
          using a second coupling pin. \label{alpha_NND}}
\begin{tabular}{ c | ccc }
   Physical coupling & $\alpha/D$ & 
   $\alpha $ (MHz)& $\Gamma^\downarrow / D$   \\
\tableline
  (0,8) & $\le 0.029$ &  $\le 0.31$ & $ \le 0.0054$ \\
  (5,3) & $0.105\pm0.008$ & $1.12\pm0.08$ & $0.07\pm0.01$ \\
  (4,4) & $0.130\pm0.007$ & $1.39\pm0.07$ & $0.11\pm0.01$ \\
  (5,8) & $0.173\pm0.006$ & $1.85\pm0.06$ & $0.19\pm0.01$ \\
  (6,8) & $0.180\pm0.006$ & $1.93\pm0.06$ & $0.20\pm0.01$ \\
  (6,8) & $0.200\pm0.006$ & $2.14\pm0.06$ & $0.25\pm0.01$ \\
\end{tabular}
\end{table}

\vspace*{-0.25cm}
\begin{figure} [hbt]
\centerline{\epsfxsize=8.6cm
\epsfbox{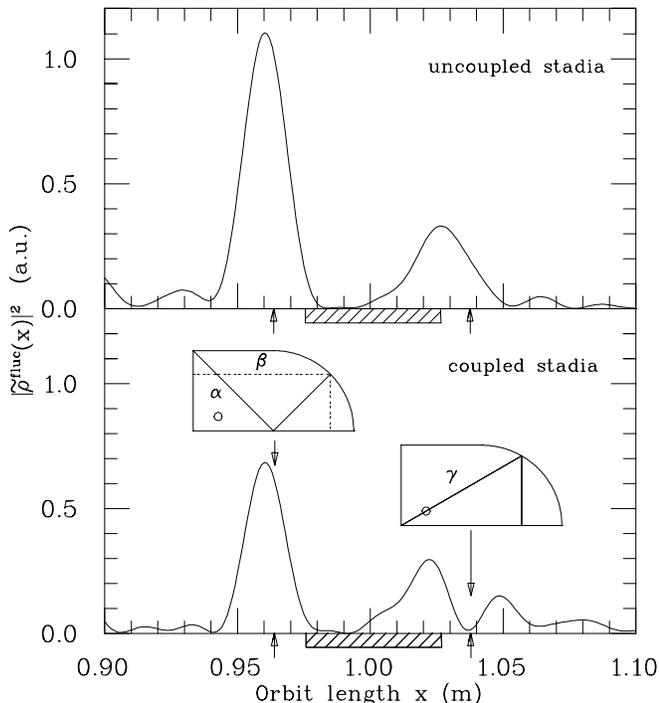}}
\caption{Fourier transforms of the spectra of the uncoupled and a coupled 
 system. The lengths of the periodic orbits $\alpha, \beta, \gamma$
 (shown in the insets) are indicated by the arrows . The lengths of the 
 whispering gallery orbits are located within the hatched area. Orbit $\gamma$
 touches the coupling pin which is marked by circles in the insets.}
 \label{FFT}
\end{figure}
\vspace*{-0.5cm}

We finally note that there is an experiment similar to the present one
performed with elastomechanical resonances in quartz blocks \cite{Ellegaard}.
Both experiments in principle allow semiclassical interpretations. The periodic
orbits for \cite{Ellegaard} are, however, very complicated. For the present
relatively simple system, we have tried to detect periodic orbits that run
back and forth through both of the coupled billiards. Systematic variations
of the integrated level density -- including those caused by the bouncing
ball orbits -- have been removed as described in \cite{Graef}. The Fourier
transform $\tilde{\varrho}^{fluc}(x)$ of the remaining fluctuating part is 
expected to display the lengths $x$ of the periodic classical orbits of the 
system \cite{Gutzwiller}. In $\tilde{\varrho}^{fluc}(x)$ obtained from the 
coupled stadia we have not been able to identify a peak introduced by the 
coupling and corresponding to an orbit running through both stadia. A very 
small part of the results is given in Fig.~\ref{FFT}. In the range of $x$ 
which is displayed, there are periodic orbits only in the $\gamma_1$-stadium. 
The shortest orbit of the $\gamma_2$-stadium is at 1.19 m. Introduction of the 
coupling changes $\tilde{\varrho}^{fluc}(x)$ at every $x$ -- whether or not
the orbits of length $\approx x$ came close to the coupling pin. This is 
expected because $|\tilde{\varrho}^{fluc}(x)|^2$ obeys a sum rule : The total 
intensity $\int dx |\tilde{\varrho}^{fluc}(x)|^2$ is given by the number of 
states. It is, however, interesting to 
see that a rather drastic change occurs in the vicinity of orbit $\gamma$. The 
peak at $x \approx 1.03$m (uncoupled case) splits into two (coupled case) and 
the interference minimum occurs at $\gamma$, i.e. the orbit $\gamma$ touches 
the coupling pin.

In summary, the dependence of the spectral statistics on the coupling between 
levels belonging to different symmetry classes has been demonstrated for a
system that simulates quantum chaos. Even subtle changes of the level 
statistics induced by small coupling parameters could be observed. The present 
experiment models mixing between any two symmetry classes.

We would like to thank O. Bohigas, G.E. Mitchell, H.-J. St\"ockmann and H.A. 
Weidenm\"uller for very helpful discussions. This work has been supported by the
Sonderforschungsbereich 185 ``Nichtlineare Dynamik'' of the Deutsche
Forschungsgemeinschaft. One of us (C.I.B.) acknowledges support by the 
Fritz Thyssen Stiftung.

\end{multicols}
\end{document}